**Graphene Oxidation: Thickness Dependent Etching and Strong Chemical Doping**

Li Liu[1,2,†], Sunmin Ryu[1,2,†], Michelle R. Tomasik[2], Elena Stolyarova[1,2], Naeyoung Jung[1,2], Mark S. Hybertsen[3], Michael L. Steigerwald[1,2], Louis E. Brus[1,2,*], George W. Flynn[1,2,*]

[1]Department of Chemistry, Columbia University, New York, NY, 10027, USA
[2]Nanoscale Science and Engineering Center, Columbia University, New York, NY, 10027, USA
[3]Center for Functional Nanomaterials, Brookhaven National Laboratory, Upton, NY, 11973, USA
[†]These authors contributed equally to this study.
[*]e-mail:leb26@columbia.edu, gwf1@columbia.edu

**ABSTRACT**

Patterned graphene shows substantial potential for applications in future molecular-scale integrated electronics. Environmental effects are a critical issue in a single layer material where every atom is on the surface. Especially intriguing is the variety of rich chemical interactions shown by molecular oxygen with aromatic molecules. We find that $O_2$ etching kinetics vary strongly with the number of graphene layers in the sample. Three-layer-thick samples show etching similar to bulk natural graphite. Single-layer graphene reacts faster and shows random etch pits in contrast to natural graphite where nucleation occurs at point defects. In addition, basal plane oxygen species strongly hole dope graphene, with a Fermi level shift of ~0.5 eV. These oxygen species partially desorb in an Ar gas flow, or under irradiation by far UV light, and readsorb again in an $O_2$ atmosphere at room temperature. This strongly doped graphene is very different than "graphene oxide" made by mineral acid attack.

**Manuscript Text**

Graphene is a zero-gap semi-metal whose electronic band structure shows linear dispersion near the charge neutral Dirac point. It exhibits novel electronic properties involving ballistic transport, massless Dirac fermions,[1] Berry's phase,[2] minimum conductivity,[3] and localization suppression.[4] Graphene strips with specific crystallographic orientations have energy gaps that increase with decreasing width.[5-7] Graphene is nearly transparent; nevertheless, Raman characterization is quite sensitive and diagnostic for crystalline quality, the number of layers, and electrical doping.[8-12] Graphene is susceptible to structural distortion,[13,14] and suspended graphene sheets show spontaneous rippling of ~1 nm.[15] Samples annealed on $SiO_2$ substrates have morphologies that depend on the substrate. AFM and STM studies have shown "flat" ~10 nm domains of 0.5 nm roughness; other regions can show bowing and bending, with deviation from a hexagonal structure.[16,17] Graphene is unusually susceptible to a Fermi level shift due to chemical doping, which can result from edge chemical functionalization in finite size pieces, and/or from charge transfer by adsorbed or bound species.

In the present study, single- and multiple-layer graphene regions are present simultaneously in a single sample (Figure1); thus, oxidation as a function of the number of layers can be directly compared. Samples were heated in an $O_2$/Ar gas flow at various temperatures (for a fixed 2 hour period, unless otherwise noted; see Supporting Information). Tens of samples, oxidized at different temperatures, were examined, and oxidative etching was observed to proceed faster in single layers than in multiple layers. Oxidation at 500°C causes etch pits 20 to 180 nm in diameter to appear in single-layer graphene, but not in double-layer sheets (Figure 1a).

Oxidative etching was carried out at 200°C, 250°C, 300°C, 400°C, 450°C, 500°C, and 600°C. AFM showed no etching of single-, double-, and triple-layer graphenes for oxidation at or below 400°C. Etch pits (~20 nm diameter) were found on single layers at 450°C (Figure S1a, Supporting Information), but not on multiple-layer samples even up to 500°C. The distributions of etch pit diameters on single graphene layers are quite broad (Figure 1; for a histogram of the pit diameters on a single-layer sheet treated at 500°C, see Figure S2a, Supporting Information). Higher temperatures induce faster oxidation. At 600°C most of the etch pits merged to give fewer, larger pits in single-layer samples. Etch pits also occurred in both double-layer and triple-layer graphene (Figure 2). Both one- and two-layer-deep etch pits were observed in double-layer graphene. In contrast, only single-layer-deep pits were observed in triple-layer (or thicker) graphene (Figure S1b, Supporting Information). Remarkably, the diameters of one-layer-deep pits on both double-layer (Figure 2a) and triple-layer sheets (Figure 2b) show narrow distributions with peaks around 220 nm (histograms in Figure S2b and Figure S2c, Supporting Information). A similar distribution was obtained in the oxidation of many-layer thick "graphite flakes" on the same substrate. This narrow distribution of pit sizes has been previously observed in the oxidation of the top layer in highly oriented pyrolytic graphite HOPG).[18, 19] The diameters of two-layer-deep etch pits on double-layer sheets are much greater (300 - 550 nm).

Oxidative etching of triple-layers is similar to oxidation of natural graphite[20, 21] and HOPG.[18, 19] Uniformly-sized one-layer-deep etch pits were observed in these studies, leading to the conclusion that oxidation was initiated at pre-existing point defects, followed by constant radial growth of the pits. This mechanism is supported by a study in which uniformly-sized pits were found on a HOPG sample after point defects were intentionally introduced by argon ion bombardment.[22] The one-layer-deep pits on our triple-layers are attributed to these same pre-existing defects since the pit diameters are nearly identical to those found on "graphite flakes" on the same substrate (Figure S1b, Supporting Information). In addition, the pit density (~ 4/μm$^2$) on the triple-layer sheet is the same as that on the graphite flake shown in Figure S1b and within the same range as that reported for naturally-occurring defect densities for various graphites.[20, 21] Thus, it appears that oxidative etching is not initiated on defect-free basal planes of triple-layers, at least at or below 600°C.

In contrast to the triple layers, we observe on single-layer graphene that $O_2$ nucleates etch pits at lower temperatures - 450~500°C, with a broad distribution of diameters (see Figure S2a, Supporting Information). A broad distribution of diameters implies continuous stochastic nucleation in time. In the $O_2$ oxidation of bulk graphite, such continuous pit nucleation only occurs at temperatures above 875°C.[23] Atomic oxygen also creates a broad distribution of pit sizes on bulk graphite[24]; here it was concluded that oxygen atom attack can nucleate and grow etch pits on a defect-free basal plane (as well as at pre-existing defects). Thus, we observe that single graphene layers are more reactive to $O_2$. This may be an intrinsic property of single layer graphene. In our experiments enhanced reactivity may additionally result from the known single layer graphene deformation by the silicon dioxide substrate.

Theoretical studies of graphene oxidation have shown lower activation barriers for attack at defects compared to attack on the pristine hexagonal basal plane.[25] As discussed previously, single-layer graphene samples annealed on $SiO_2$ substrates can be structurally deformed showing regions of curvature (sometimes conformal to the rugged substrate) and domains of lower than hexagonal symmetry.[16, 17] These curved regions, both long wavelength ripples and local bonding distortions, should result in some $sp^3$ C orbital character and π-orbital misalignment. This is expected to lead to significantly increased reactivity[26-28] as has been observed in $O_2$ oxidation of carbon nanotubes.[29, 30] Also, $Cs^+$ ions trapped below the top layer in bulk graphite can nucleate oxidation[31]; thus, in the present study surface oxide charges may also be capable of nucleating etch pit growth in single-layer graphene.

Double-layer graphene sheets are intermediate between single layers and bulk graphite in their etching behavior. They show two-layer-deep etch pits with a broad size distribution, as well as mono-dispersed one-layer-deep pits. While the latter are attributable to pre-existing defects, the origin of the former may be substrate-induced nucleation. In this connection, free standing double-layer graphene[15] does show spontaneous deformation, albeit with lower amplitude intrinsic ripples than single layers. Note that the average size of two-layer-deep pits in the present study is larger than that of one-layer-deep ones. Multiple-layer-deep etching on bulk graphite is well documented to proceed faster than monolayer deep etching.[22, 32] This effect was also observed in the present double-layer samples. Multiple-layer-deep defects have also been observed to be a prerequisite to nucleate multiple-layer etching.[22, 32] Since multiple-layer-deep defects rarely appear in graphite,[32] the two-layer-deep pits on double-layers may be attributed to substrate-induced nucleation. Triple-layer graphene, however, does not show any significant difference from graphite flakes in terms of oxidation pit size and distribution. Thus, triple or thicker layer graphene can be considered as bulk graphite with respect to oxidative etching.

Raman scattering can be employed to directly probe oxidized graphitic material. The shape of the D* mode (~2680 $cm^{-1}$)[9, 11, 12] and the relative intensities of the doubly degenerate G mode (~1580 $cm^{-1}$) have been used to establish the number of layers present in graphene samples.[9-12] In unoxidized, single-layer graphene the D disorder band is barely detectable ($I_D/I_G \leq 0.01$, Figure 3a). The intrinsic defect concentration is very low. The D mode grows in as oxidation proceeds for all layer thicknesses. There is an empirical correlation between the intensity ratio of the D mode to that of the G mode ($I_D/I_G$), and the average domain size ($L_a$).[33-36] This ratio is plotted in Figure 3b for single-,

double-, and triple-layers. Triple-layers show significant D mode activity only above 500°C. Double-layers show a higher $I_D/I_G$ than triple-layers when oxidized at 600°C. In single-layer samples, $I_D/I_G$ increases slightly for oxidation at 300°C. When large etch pits grow in at higher temperature, the $I_D/I_G$ empirical relationship[34] yields a domain size $L_a$ far smaller than the typical spacing between visible etch pits. This implies that the D intensity is due not only to the visible pits, but must also reflect local disorder induced by oxidation. These changes occur at lower temperature on single-layers than on double- and triple-layers.

The G mode position before oxidation is 1583±2 cm$^{-1}$, which implies that the initial graphene (measured in air) is nearly intrinsic. Both the G and D* modes shift to significantly higher frequency with oxidation (Figure 4a and Figure 3a). Most of this shift occurs between 200°C and 300°C irrespective of thickness (Figure 4b), unlike the D band growth and pit etching that appear at higher temperature. These shifted spectra appear to result from a different, lower temperature process. These relatively narrow spectra are very different than those of amorphous "graphene oxide" made by reaction with strong mineral acid[37-39]. The G band width of graphene oxidized at 300°C is 9±2 cm$^{-1}$, while that of "graphene oxide" varies in the range of 80~110 cm$^{-1}$. They are also quite different than the predicted Raman spectra of graphene with covalently bonded OH or bridging epoxide O atoms on sp$^3$ C atoms.[39] However, the spectra are quite similar to those of strongly hole doped graphene in electrostatic gate devices.[40] We conclude that $O_2$ oxidation at 200°C to 300°C creates strong hole doping in graphene. In this temperature range, the very weak D band implies that there are very few sp$^3$ C atoms in the graphene. The magnitude of the shift is largest for single-layer samples and decreases with increasing thickness. The double-layer G mode for 600°C oxidation has a lower-energy shoulder (Figure 3a), implying asymmetry between the two graphene layers, consistent with the AFM images.

The doping level is very high. The maximum G-mode shift for single layers (~23 cm$^{-1}$) at higher temperatures has only been achieved in an electrochemical top-gating configuration.[8, 40, 41] From this top-gate experiment we estimate a hole density ($n$) of 2.3x10$^{13}$ cm$^{-2}$ with a corresponding Fermi energy ($E_F$) shift of ~0.56 eV below the neutrality point (Figure 4, inset). The maximum charge density corresponds to one hole per ~170 carbon atoms (approximately one hole per 4.5 nm$^2$ area).

As discussed earlier, the initial graphene in air before oxidation is not doped by physisorbed $O_2$ or water. Also, graphene devices on silicon dioxide substrates are experimentally intrinsic after heating for long periods in high vacuum at 150°C.[42] Thus direct physical contact with silicon dioxide by itself does not dope graphene. Our samples are doped by electron transfer from graphene to oxygen species, which may additionally react with water. Perhaps this species is a basal plane bound hydroperoxide or endoperoxide, such as is also thought to occur in carbon nanotubes.[43] In addition, it might be an uncharacterized charge-transfer complex between a nearby electronegative $O_2$ and electron rich graphene.[44] Note that adsorbed $O_2$ on silicon dioxide is a known electron acceptor.[45] Such a complex, if present, would produce a new charge-transfer electronic absorption band.

To further explore doping, the Raman spectra of graphene oxidized at 300°C was subsequently obtained under both Ar and $O_2$ flow at room temperature. Remarkably, the G mode, which had been upshifted by ~18 cm$^{-1}$ by the 300°C oxidation, downshifts by 3.5 cm$^{-1}$ in Ar, and then upshifts to its original position when treated again with $O_2$ (Figure 5). We also found that far UV Hg lamp irradiation under an Ar flow significantly downshifts the G mode, decreasing the hole density by ~70%. More surprisingly, the G mode upshifts again to the "as-oxidized" value when exposed to $O_2$ flow at room temperature. A double-layer sheet oxidized at 300°C also showed similar behavior. Thus doping is mainly caused by oxygen species bound to graphene, some of which are in equilibrium with gaseous $O_2$ at room temperature. Analogous $NO_2$ gas-surface equilibria and UV-induced desorption have been observed in high vacuum graphene transport experiments.[42] Similar $O_2$ equilibria have been seen in carbon nanotubes,[43, 46-48] pentacene transistors,[49] and generally in aromatic molecules.[50] While taking Raman spectra of oxidized graphenes, the laser intensity was kept very low to minimize desorption and/or further oxidation due to laser-induced heating.

We found that simply heating a fresh graphene sample at 300~400°C in forming gas (10% $H_2$ and 90% $N_2$) or heating it in ultra high vacuum (400°C) (without subsequent oxidation) produces hole doping when the sample is exposed to the atmosphere. This doping was two-thirds (sometimes more) of that obtained for heating the sample in oxygen. It may be that graphene is doped by $O_2$ released from the silicon oxide, or that heating structurally deforms ("activates") graphene allowing it to react with atmospheric $O_2$ when removed from the oven (or from the UHV environment) for Raman characterization. (Note that silicon oxide has been used as a capping layer to control oxidation of multilayer graphene.[51]) Ground state triplet $O_2$ interacts weakly with flat graphene.[52, 53] Higher temperature reaction may be initiated by excited singlet $O_2$, yielding the endoperoxide or hydroperoxide adduct.[43, 54] However, the facile decrease and recovery of doping in Figure 5 suggests that heated graphene is subsequently reactive to ground state $O_2$ in air at room temperature. This supports the heating induced graphene activation mechanism. As previously discussed, graphene annealed on atomically rough $SiO_2$ substrates significantly deviates from the ideal flat hexagon structure in STM images.[16, 17] Such deformation will increase graphene reactivity. For example, the strongly strained aromatic molecule helianthrene binds ground state oxygen as an endoperoxides.[55] Generally, UV light easily decomposes endoperoxides into $O_2$ and the original aromatic molecules[50] which is consistent with our UV-induced undoping. It is remarkable that the strongly hole doped graphene with basal plane adsorbed oxygen species does not show a significant D band.

Even at higher temperatures when etch pits grow, this basal plane charge transfer process is responsible for high doping. In the oxidation temperature range of 450~600°C, there are typically 20~30 etch pits per square micron on a single-layer sheet. At 450°C, for example, the pit diameter is ~ 20 nm. Thus the single-layer sheet has a density of edge carbons of 6200~7400 $C_{edge}$/$\mu m^2$ (a zigzag (armchair) edge has 4000 (4700) $C_{edge}$/$\mu m$). Even assuming one electronic hole coming from each oxidized $C_{edge}$, the resulting charge density is far smaller than the $2.2 \times 10^5$ charges/$\mu m^2$ deduced from the G-mode upshift at this temperature.

At higher temperatures irreversible etching occurs. The extent of irreversible oxidation varies with layer-thickness: single-layers oxidize more quickly than double-layers, which oxidize more quickly than triple-layers and bulk graphite. The thickness dependence should be linked to the geometries of the oxidation transition states and intermediates. Any covalent bond formed between an oxygen atom and a carbon atom in graphene will result in one or perhaps two tetrahedral carbon atoms. This implies an abrupt and local bend in the carbon sheet. The height of the oxidation activation barrier is directly related to the energetic cost of this bend: the stiffer the carbon sheet, the higher the activation barrier. There is a strong conformal attraction between graphene sheets. Thus, although sheets can easily slip across one another, perpendicular motion is very costly. Any bending of a single sheet in a graphene stack will be opposed by the inter-sheet attraction. The graphene stack acts like a leaf spring; the larger the number of sheets, the stiffer the "spring" and, therefore, the higher the activation barrier.

In summary, this study has demonstrated that single- and double-layer graphenes differ significantly from one another in oxidative etching, and that both differ from triple-layers which behave much like bulk graphite. Low temperature (~300°C) reaction with $O_2$ creates strongly hole doped graphene without etch pit nucleation, and with a negligible D disorder Raman band. The electron acceptor might be an endoperoxide, a hydroperoxide, and/or a charge transfer complex, bound on the basal plane. Heating graphene on a silicon dioxide surface in either a reducing atmosphere or in ultra high vacuum activates the sample for a subsequent hole doping reaction under room temperature atmospheric conditions. Since graphene materials and devices are typically supported and annealed on solid substrates,[56] substrate-induced effects, such as those observed in the present experiments, are likely, and in fact are a critical aspect of graphene materials science.

**Acknowledgements**
This work was funded by the Department of Energy under Grant Nos. DE-FG02-88ER13937 (G.W.F.), DE-FG02-98ER14861 (L.E.B.) and DE-AC02-98CH10886 (M.S.H.). We acknowledge financial support from the Nanoscale Science and Engineering Initiative of the National Science Foundation under NSF Award Number CHE-06-41523 and by the New York State Office of Science, Technology, and Academic Research (NYSTAR). Equipment support provided by the National Science Foundation under grant CHE-03-52582 (G.W.F.).

**Supporting Information Available**
Description of experimental methods, AFM and Raman spectroscopy studies on adhesive-free graphene samples, and histograms of etch pit diameters in single-, double- and triple-layer graphene sheets. This material is available free of charge via the Internet at http://pubs.acs.org.

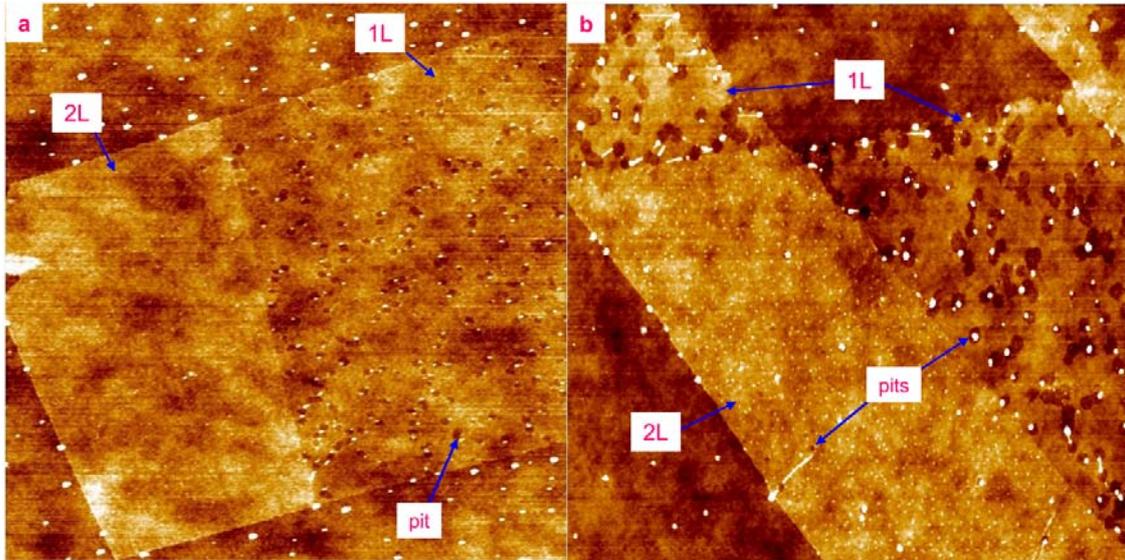

**Figure 1.** AFM images of oxidized single-layer (1L) and double-layer (2L) graphene. (a) (5.06 x 5.06 μm$^2$) oxidized at 500$^\text{o}$C for 2 hours ($P$(O$_2$) = 350 torr). (b) (6.95 x 6.95 μm$^2$) oxidized in reduced O$_2$ pressure at 600$^\text{o}$C for 40 minutes ($P$(O$_2$) = 260 torr).

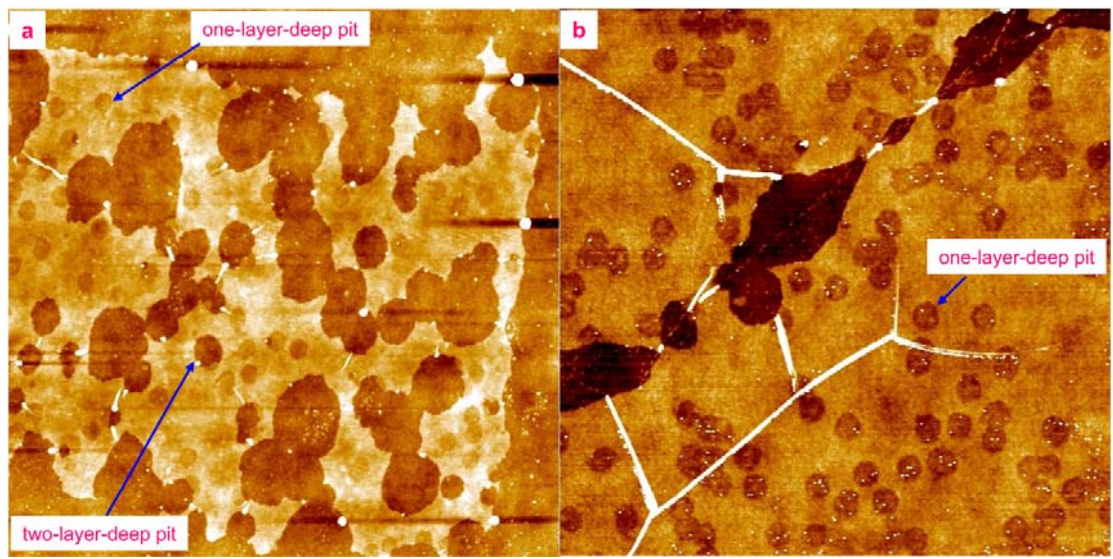

**Figure 2.** AFM images of oxidized double-layer and triple-layer graphenes. (a) double-layer graphene oxidized at 600°C for 2 hours (7.00 x 7.00 μm²). (b) triple-layer graphene oxidized at 600°C for 2 hours (5.00 x 5.00 μm²). Both one-layer- and two-layer-deep pits appeared on the double layer, whereas only one-layer-deep pits could be found on the triple layer. The gaps diagonally crossing (a) were probably due to pre-existing structural line defects.

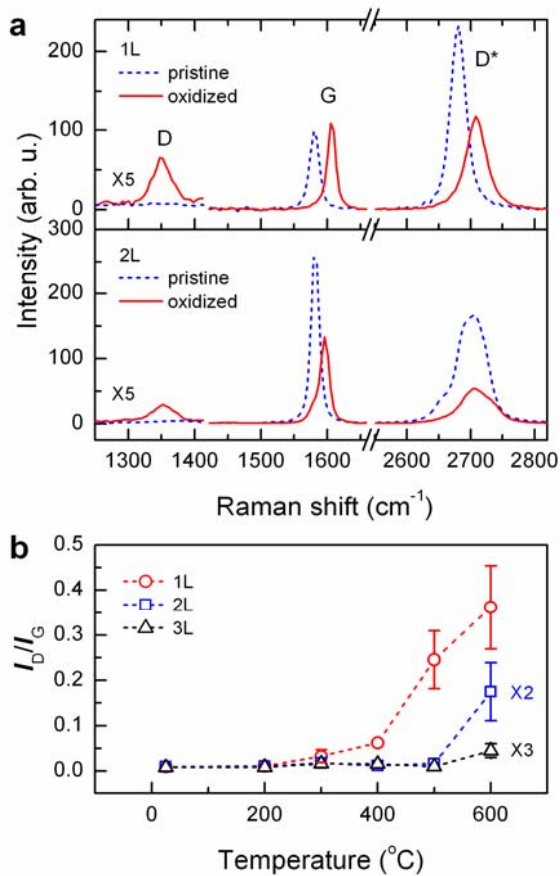

**Figure 3.** Raman spectra of pristine and oxidized graphene. (a) (Upper) Raman spectra of single-layer (1L) graphene: pristine (dotted), and oxidized at 500°C for 2 hours ($P(O_2)$ = 350 torr) (solid). (Lower) Raman spectra of double-layer (2L) graphene: pristine (dotted), and oxidized at 600°C for 2 hours ($P(O_2)$ = 350 torr) (solid). Spectra near D-mode were enlarged for clarity. (b) D-mode to G-mode integrated intensity ratio ($I_D/I_G$) as a function of oxidation temperature: single-layer (1L, circle), double-layer (2L, square), and triple-layer (3L, triangle) graphene.

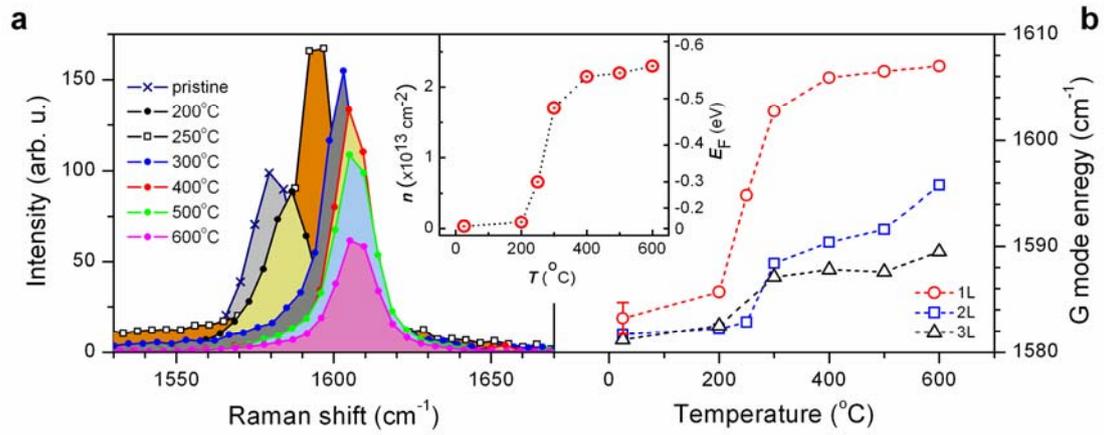

**Figure 4.** Raman G-mode and chemical doping of oxidized graphene. (a) Raman spectra near the G-mode of single-layer graphene samples oxidized at various temperatures for 2 hours ($P(O_2)$ = 350 torr). (b) Peak positions of the G-mode as a function of oxidation temperature: single-layer (1L, circle), double-layer (2L, square), and triple-layer (3L, triangle) graphene. The error bar for the pristine 1L graphene (at 23°C) represents the standard deviation of 10 samples. The inset shows the hole density (***n***) of oxidized 1L graphene estimated based on Ref. 40 and the Fermi energy ($E_F$) calculated from $\boldsymbol{E}_F = -h\boldsymbol{v}_F(\boldsymbol{n}/4\pi)^{1/2}$, where h and $\boldsymbol{v}_F$ are the Planck constant and Fermi velocity ($10^6$ m/s), respectively.

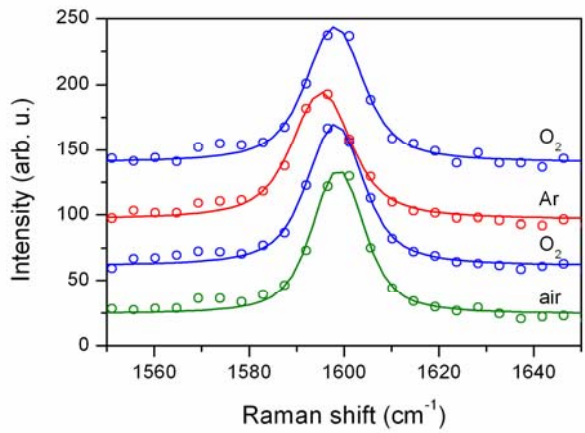

**Figure 5.** Raman G-mode of oxidized single-layer graphene in various ambient gases. The Raman spectra were taken for one sample, oxidized at 300°C for 2 hours ($P(O_2)$ = 350 torr), consecutively in air, $O_2$, Ar, and $O_2$ flow. All spectra were taken at an identical spot to avoid spatially dependent spectral variations. Solid lines are Voigt fits to the experimental data (circles); the instrument spectral resolution was 8.4 cm$^{-1}$.

**Synopsis Graphic**

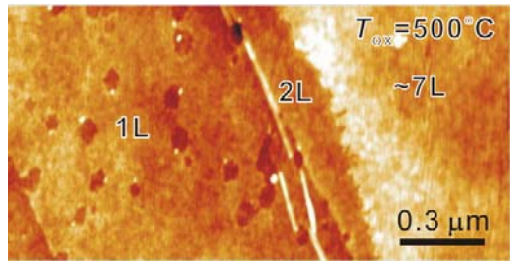